# Experiential Learning Styles and Neurocognitive Phenomics

## Włodzisław Duch
*Nicolaus Copernicus University*[1]

### ABSTRACT

Phenomics is concerned with detailed description of all aspects of organisms, from their physical foundations at genetic, molecular and cellular level, to behavioural and psychological traits. *Neuropsychiatric phenomics,* endorsed by NIMH, provides such broad perspective to understand mental disorders. It is clear that learning sciences also need similar approach that will integrate efforts to understand cognitive processes from the perspective of the brain development, in temporal, spatial, psychological and social aspects. The brain is a substrate shaped by genetic, epigenetic, cellular and environmental factors including education, individual experiences and personal history, culture, social milieu. Learning sciences should thus be based on the foundation of *neurocognitive phenomics*. A brief review of selected aspects of such approach is presented, outlining new research directions. Central, peripheral and motor processes in the brain are linked to David Kolb inventory of the learning styles. Even with such simplified model based on connectome for 3 regions 84 brain states may be distinguished. Transitions between such states are quite likely, so it is not clear that human behavior, including learning styles, can be clustered into a small number of meaningful types.

### KEYWORDS

Learning sciences, brain, connectome, phenomics, neuroeducation, genotype, phenotype, memory, learning, cognition, learning styles/preferences.

### INTRODUCTION

Successful learning requires formation of stable memory patterns that can be accessed by future brain processes. Brain substrate, the incredibly complex network of neurons, glia cells, genetic and epigenetic biochemical processes, maintains a delicate balance between stability and plasticity, the need to remember past experiences and the need to learn new ones. Full understanding of the biological learning processes requires explanations of processes on many levels of complexity, from genetic and molecular level to behavioral levels. The word *phenomics* has traditionally been concerned with the study of phenotypes, measureable traits of an organism. Rapid development of many branches of biology has broaden the phenomics perspective. In analogy to the concepts of *genome* and *proteome* that signify all genes and proteins *phenome* is the set of all phenotypes of cell, tissue, organs, or the whole species, expressed by their measureable traits. The *Human Phenome Project* (Freimer & Sabatti 2003) is gaining momentum, collecting all kinds of information about phenotypes and linking it with genomic and any other available information that may be useful to understand many aspects of human behavior.

[1] Włodzisław Duch, Department of Informatics, Faculty of Physics, Astronomy and Informatics, and Neurocognitive Laboratory, Center for Modern Interdisciplinary Technologies, Nicolaus Copernicus University, Toruń, Poland. wduch@umk.pl

In 2008 the *Consortium for Neuropsychiatric Phenomics* (CNP) has been formed in the USA to focus on diagnostic and therapeutic methods in various mental disease, trying to understand biological mechanisms underlying phenotypes. Initial project will provide detail characteristics of groups of 300 people, each group suffering from Schizophrenia, Bipolar Disorder, and Attention-Deficit/Hyperactivity Disorder (ADHD), comparing genetics, brain structures, brain functions (memory, response inhibition) and behavior with 2000 healthy people. This project should help to understand how disruptions of normal functions influence phenotypes at each level (cf. Research Domain Criteria (RDoC) of National Institute of Mental Health, Insel et al., 2010). This is obviously only the first step towards understanding of mental disease, and the progress so far has been slow but steady. Unravelling brain mechanisms and integrating information about the impact of environment on genetic and epigenetics mechanisms, formation of proteins that build cellular systems, development of neurons and neural systems, creation of signaling pathways, formation of cognitive phenotypes, manifestation of symptoms and syndromes in different contexts is a great challenge, but there is simply no alternative if we want to understand such processes in details.

Comparable framework for learning sciences has not yet been formulated. **Learning sciences** is an interdisciplinary field studying human learning, focused on psychological levels related to cognitive, social and cultural factors that influence learning process and help to design better learning environments. Understanding learning processes, and placing life-long education on solid scientific foundations, requires comprehensive approach that should cover all aspects of phenomics. This is true not only for special needs education. The problem is even more difficult than in case of neuropsychiatric phenomics, as the goal is not just repairing disrupted function, but unfolding full human potential and improving quality of life. In this short article the idea of ***neurocognitive phenomics*** is proposed, a new transdisciplinary direction that will take decades to complete. At present learning sciences is mostly focused on psychological-cognitive, social and cultural factors, although several doctoral programs in this field include neuro-cognitive aspects. Educational neuroscience develops in parallel and there seems to be a little overlap with learning sciences. Complete understanding of all processes that contribute to learning cannot be done treating cognitive systems as black boxes. Artificial intelligence has tried it for more than half a century without much success. Recent progress was achieved by paying attention to the neural level. However, not all phenomic levels may be equally important. As an example explanation of experiential learning styles (Kolb 2015) is presented at relatively high level of information processing in the brain, disregarding deeper phenomic levels.

In the next section various levels of phenotype description are presented in relation to learning, and in the third section experiential learning styles and their relation to the infromation flow between different areas of the brain are discussed. A brief discussion closes this paper.

FROM GENES TO LEARNING STYLES

Many branches of science contribute to understanding of developmental and behavioral processes. In Fig. 1 processes at seven levels of hierarchical description are presented. Any division of biological processes into such levels is a result of specific research approaches, different views that help to describe undivided reality. Genetics, neurophysiology, behavioral methods and psychology provide complementary descriptions of the learning processes that should be integrated into one comprehensive theory. Levels that are higher in the hierarchy may have their own emergent properties that do not make much sense at the lower level. Processes at low levels contribute to our understanding of higher levels, from genetics to cognition, influencing processes at all stages and changing individual phenotypes. Higher-level processes provide constraints for description of the lower level ones. Simple linear causality is not applicable here; educational interventions may change phenotypes at lower levels, depending on their current state. Psychiatry and special education needs aim at identification of serious distortions of this process. Learning sciences should also be concerned with more subtle effects. The major factors that influence brain development include genetics, nutrition, infection, immune system responses to environment, traumatic experiences, influence of family and society, and of course education.

GENES AND PROTEINS

Biological systems have incredible complexity. In a single cell more than half of its dry mass is due to proteins. There are now more than 16 million known protein sequences, and about 85,000 full structures in the Protein Data Bank (PDB). At the molecular level basic building blocks of cells are constructed according to the master plan that is stored in the DNA. Millions of proteins are formed using information in RNA and DNA, constructing all cells, including neurons, in an environment that needs to supply many types of elements (about 60 types have been found in human body) in sufficient quantities to ensure normal development. Serious intellectual deficiency (IQ less than 70) affects 1-3% of general population and has been linked to mutations in a few thousand genes.

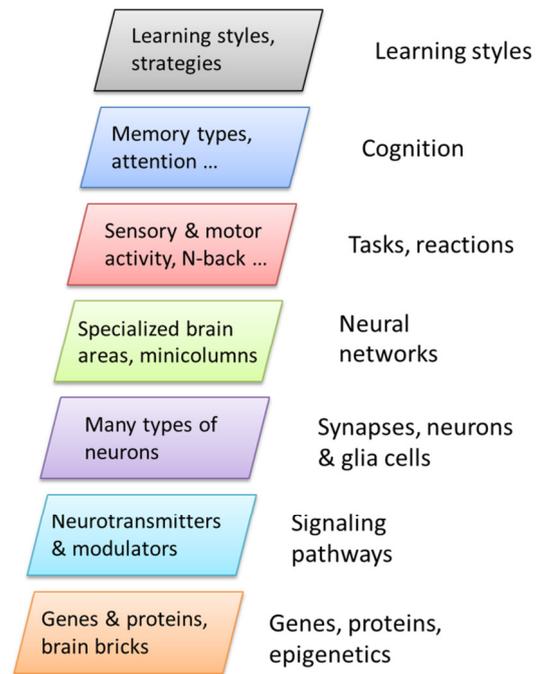

The execution of the DNA building plan is controlled by the environment through many types of epigenetic processes, altering gene expressions without changing the DNA sequence. Many types of epigenetic processes have been discovered and probably many more exist. DNA methylation (adding $CH_3$ molecules to predominantly cytosine bases) has been studied in relation to serious health conditions such as cancer for three decades. The *DNA Methylation Society* has its own journal *Epigenetics*, covering medical, behavioral, psychological and nutritional epigenetic effects. Modification of chromatin (complex of DNA and proteins found in cell nucleus) may occur through methylation, microRNA or enzymes, and may silence one or both genes. This process, called in genetics *imprinting*, has also been implicated in many diseases. Moreover, epigenetic changes may be inherited without modification of genetic sequence.

Development of individual traits starts already in prenatal period. Many traits are inherited and shape the structure of the whole organism. In the womb basic personality traits, such as responsiveness and temperament, are already determined to a large degree (Pesonen et al. 2008). The length of pregnancy may have strong influence on the development, cognitive abilities and tendency to acquire major disease. Epigenetic effects occur through all life but are strongest in the womb. Food intake, vitamins, drugs, alcohol, inflammation, viruses and bacteria causing disease, exposition to various pathogens in the environment are some of the factors that may influence epigenetic processes, leading to metabolic problems, blood defects and degeneration of internal organs, as well as more subtle influences on the brain. Phenotypes of identical twins raised in different conditions have shown different patterns in muscle, lymphocyte and other tissues. This is an area of intensive research, focused around the *Human Epigenome Project* initiated in Europe in 2003. Although this project is oriented towards cancer and other serious disease implication in child development, neurological and psychiatric disease and aging has also been pointed out. Epigenetic processes differ in various tissues, so this project will be more complicated than the *Human Genome Project*.

Epigenome may be affected in a simpler way than genome itself, therefore it should be an easier target to control. Immediate goal of this research is to prevent abnormal development, but long-term goal may be to ensure optimal conditions for development of the brain and the whole organism.

<div align="center">SIGNALLING PATHWAYS</div>

Complexity of organisms is not directly related to the number of genes. Humans have probably about 20,000 genes coding proteins (and about the same number of "RNA genes"). Nematode worm has similar number of genes, and wheat has about 4 times as many as humans. What seems to be more important is the complexity of genetic regulation, epigenetic and posttranslational processes, and interactions among proteins. This interaction network, called *interactome*, in case of humans has been estimated to be about 650,000. It is several times larger than in simpler organisms (Stumpf et al. 2008). Proteins interact with other molecules creating functional complexes and networks of interactions that form metabolic pathways. Inside the cell they regulate concentration of biochemicals in different cell structures. Through various pores and channels in cell membranes substances are exchanged, bringing nutrients to the cell and allowing for communication with other cells. Some molecules that are part of membrane receptors are sensitive to the external environment, and they influence processes within the cell. The development and well-being of the organism critically depends on all signalling pathways.

Neural stem cells at different locations mature into different types of neurons and glia cells depending on the density of *neurotrophins*, glycoproteins such as reelin, extracellular matrix, retinoic acid (vitamin A metabolite), cerebrospinal fluid and the vascular environment. Many specific pathways that facilitate, inhibit or pattern neurogenesis have been discovered, but this is still largely uncharted territory. In adult brains of mammals neurogenesis is limited to hippocampus and olfactory bulb areas. This helps to ensure stability of memory and behavior that has already been learned. Fundamental questions about cellular and molecular pathways still need to be answered before control of these processes and therapies for neurodegenerative disease will be possible.

Inspired by genomics, hundreds of *"omics"* fields of study in the life sciences have been created (listed at omics.org). Environmental factors that organisms are exposed to may cause disease in prenatal as well as postnatal stages of development. Research on the effects of such factors has been called *exposomics*. Investigation of nutritional aspects of bioactivity is called *foodomics*. So far attention is paid more to the dangers than to the possible benefits. Monitoring of environmental and nutrition effects on signalling pathways will contribute to preven-tion of developmental problems, reducing the number of children who need special education. Observations of factors that influence behavior and induce physiological changes, called *ethomics*, started with creation of "ethoscopes", machines that perform real-time tracking and profiling of behavior by using machine learning algorithms (Geissmann et al, 2017). This approach has been now extended to *human ethomics*, measuring many physical parameters characterizing human behavior in specific conditions, like extreme car driving (Rito Lima et al, 2020). We should expect similar developments in learning sciences.

<div style="text-align:center">NEURONS</div>

The brain is not a general purpose computer. It is rather a multi-agent system, a highly specialized device that offers a large number of automatic responses, many of which can be adapted as a result of learning to realize complex cognitive functions. Cells in the brain – neurons and glia cells – are quite unique; they are the longest living cells in the body. Neurons come in a large variety, differing in size, morphology, length of axons and number of dendrites, dendritic spines, types of synapses and receptors. Some have very long axons providing links between distant brain areas (transversal connections, fascicle bundles, association fibres). Information is distributed to all parts of the brain, synchronizing their activations needed to solve complex problems. The brain is connected to distant parts of the body through efferent and afferent nerve fibers. Many types of neurons have short connections. They regulate the level of local arousal, keeping the excitation at low level. A rough estimate of the number of different types of neurons is based on 1000 grey matter regions, each containing on average about 10 unique types of neurons, giving about 10000 types of neurons. Recent estimations of the number of neurons in the human brain (Lent et al. 2012) converge at 86±8 bln neurons in the whole brain, but only 16±2 bln in the cortex, less than 2 billion in the subcortical structures, the spine and other parts of the body, and about 69±7 bln in cerebellum, involved mostly in motor control and motor learning. This shows how important the precise control of movement is. Perception, memory and cognitive processes serve to control goal-directed action.

The development of neurons may go wrong in many ways, caused by a large number of factors: the whole brain may develop in a wrong way at a gross anatomic level, genetic mutations, environmental and epigenetic factors may lead to misfolding of proteins damaging neural internal structures (mitochondria and other organelles), membranes, dendrites, spines, synapses. Neurotrophic factors that control proliferation and migration of neurons, neural apoptosis (programmed cell death) and synaptic pruning that should leave only useful functional connections between neurons may fail due to a number of reasons. Neurons need to send spikes to stay alive, and may send bursts of no more than a few hundred spikes per second before they take a break and move into slow spiking mode.

Problems with development of neurons and their migration to proper places may lead to death of the fetus or to a serious neurodegenerative and psychiatric disease after birth. At present not much can be done about it. Because so many factors may be responsible for neural dysfunctions it is not surprising that in extensive studies of genomes of autistic people mutations in hundreds of genes have been weakly correlated with about 20% of autism cases (for the remaining 80% correlations are too weak to be statistically significant). More than half of these mutations are also correlated with other mental disease. It is quite likely that serious mental disease such as autism or schizophrenia, and neurological disease such as epilepsy, result from subtle dysfunctions of neurons that make them incapable of forming proper networks. Such disease may have infinite number of variants, because the damage of different severity may occur in one or many types of neurons, in one or many brain areas, axons and dendrites may not branch correctly, brain circuits may be miswired. Each case of brain disease may be unique and manifest itself in slightly different way.

NEURAL NETWORKS

Neurons are densely connected, with 1-2 bln synapses in one cubic millimeter of the neocortex. The total number of connections is of the order of $10^{14}$, or 100 trillion, most of these connections are between neurons in the neocortex. In the first and the second year of life several millions of new synaptic connections are formed every second. The total number of connections grows after birth, reaching its peak around the third year of life. The process of pruning and creating new connections continues for about 20 years, and it is a bit different in the left and right hemisphere. Learning is largely based on specialization, efficient processing of information, specific associations that have to be established, and that requires better structural integration of brain networks. During adolescence density of gray matter still decreases, short-range connections are pruned and long-range myelination helps to synchronize distant brain areas. Repetition creates stronger pathways between neurons involved in representation of concepts that are learned, making associations more automatic. This is important in early perception, sharpening discrimination of sensory stimuli, seeing textures, contrasts and shapes, tactile sensations and auditory discrimination, filtering speech sounds that facilitate understanding language in a speaker-independent way. All sensory systems have to quantize the incoming signals to enable invariant recognition. Elementary patterns in speech, general sounds, visual and other stimuli are extracted in the primary sensory cortices. Complex patterns are built from these basic cortical activations by secondary and higher-order cortices. Brains of infants and babies develop these abilities in a spontaneous way in a proper environment without direct supervision or monitoring (Lamb et al. 2002). Positive effect of rich environments on formation of neural networks has been documented in animal research.

The complexity of the meso and micro-scale is overwhelming and cannot be precisely controlled by genetic processes. Neurons in the neocortex are organized in 6 layers and in cortical columns, with diameter of a fraction of millimeter. Each column contains a few tens of thousands of neurons of many types, forming densely connected vertical microcircuits, connected horizontally in a less dense way. The cortex is composed from a few millions of such minicolumns, and they serve as computational units with rich internal dynamics. At the macroscale brains have similar structures. It is surprising, but theoretical analysis shows that local microcircuits may have random connectivity and huge diversity to allow for more robust and efficient analysis of incoming spatiotemporal signals (Maass et al, 2002). "Small world" structure of brain networks, with important hubs where many connections converge, is also contributing to the efficient information processing.

SYSTEM LEVEL

At the system level the whole brain is analyzed focusing on regions of interest (ROI), anatomically and functionally specialized brain areas. The *Human Connectome Project* started in the USA in 2009, mapping connectivity between about 1000 regions of interest to create "network map" for healthy adults and people with various mental problems. This project has been followed by plans to make connectomes at more detailed spatial scales, at mesoscopic and microscopic levels. Precise maps of connectomes of primitive organism (such as nematodes) showing connections and functional activity for single neurons have been created. In time domain development of connectomes at the prenatal period and in early infancy (*Developing Human Connectome Project*) has been started in 2013. The Lifespan Connectome consortium started at the same time, collecting data on babies (age 0-5), children (age 5-21) and investigating aging processes (age 36-100). Another direction was taken by the "*Connectomes Related to Human Disease*" project, sponsored by several USA national insti-tutes working on addictions, aging, psychiatric and neurological disorders, and eye diseases. Various mental problems are due to impaired information flow in the brain, and can be recognized using properties of connectomes: anxiety and depression in adolescents, frontotemporal dementia, psychosis and many other problems. Connections in the brains of people suffering from Alzheimer's disease, autism or schizophrenia clearly differ from those in healthy brains. Connectomes for thousands of people have now been analyzed. Features of individual connectomes can explain not only brain disorders, but also factors relating intelligence to flow of information in the brain (Beaty et al, 2018; Sunavsky & Poppenk, 2020).

Brain structures are not directly connected to single cognitive function. Each brain region may contribute to many functions, and vice versa, each function may be realized engaging different subsets of brain regions (Anderson, 2010). Only a partial localization of functions is possible. There are significant individual differences in the way brains processes information, and they certainly influence learning style of individuals. These differences are relatively subtle comparing to the changes observed in brains of people with serious mental problems. Pharmacological interventions target large subsystems that use the same type of signaling pathways based on various types of neurotransmitters and their receptors, for example influencing the level of serotonin or dopamine. Such interventions are very coarse, changing dynamics of the whole brain in many ways. Some neurotransmitters are produced locally in neurons all around the brain, but most are created by specific nuclei in the brain stem, and transported to large areas of the cortex, including serotonin (raphe nuclei), norepinephrine (locus ceruleus), or dopamine (ventral tegmentum area and substantia nigra in the midbrain).

Brain stem is responsible for homeostasis, many automatic responses, selection of the global behavioral state, activating and inhibiting large brain areas depending on the overall context, with more precise selection of actions done by big subcortical nuclei called basal ganglia. Reward information is used to choose, learn, prepare and execute goal-directed behavior, mediated by dopamine neurotransmitters, engaging medial temporal cortex involved in the detection and prediction of rewards. Orbitofrontal cortex and amygdala evaluate relative reward values and expectations (Berns, 2005; Gottfried, O'Doherty & Dolan, 2003; Schultz, 2000). The overall level of cortex arousal determines the awareness level, and is controlled by the reticular formation in the brain stem.

Representation of goals is maintained in parietal, premotor and dorsolateral prefrontal cortex (Berns, 2005). Motivation, resulting from anticipation of rewards or conditioned positive emotions, is correlated with activity of ventral striatum. Executive functions, such as planning, reasoning, abstraction, initiation and disinhibition of behaviors, are strongly correlated with results of verbal and visual memory tests and with the IQ tests (Duff, et al. 2005). Short-term memory is based on arousal of subsets of neurons that change the flow of neural activation, therefore it has very limited capacity, but traces of memory activation (neurons that were active are easier to excite) may prime related activity and influence decisions long after the active memory content has changed. Holding several things in working memory, and being able to store and quickly return to previous thoughts, is very important for problem solving, understanding of complex situations and use of language. The more parallel plans the brain is able to pursue the better. Prefrontal cortex is the key brain area performing working memory functions, with capacity for only 3-4 visual objects (Baddeley, 2002), and about 7±2 chunks of simpler information, such as digits or random letters (Cowan, 2005). A distinct verbal working memory system is used by the brain to analyze the syntactic structure of a sentence and determine its meaning (Caplan & Waters, 1999). Higher capacity of the working memory is strongly correlated with intelligence.

COGNITION LEVEL

Cognition level contains mental processes that can be investigated externally, using behavioral tests. These processes are described in psychological and information processing terms and have only recently been partially linked to the brain processes. Many cognitive functions involving perception and motor control have now neural network models that can be verified using computational simulations and neuroimaging. Thoughts arise when large-scale synchronization (binding) of different brain areas is created, creating global states that are clearly distinguished from noise resulting from random excitations. Sudden strong local activity (as in specific language impairment, or epilepsy) may disperse thoughts, making it difficult to focus and learn. Deficits in quantization of basic speech sounds (phonemes) lead to phonological dyslexia and learning problems at school.

The brain has rich neural structure, with many synaptic connections, therefore it may adapt to the complex challenges posed by the environment and some internal malfunctions. The density of synaptic connections has some influence on the sensory event-related potentials (ERPs), one of the oldest techniques used in brain research, based on electroencephalography (EEG). For example auditory cortex reaction to simple repetitive stimuli (speech or general sounds), measured by averaging EEG signals from the temporal brain areas. This signal shows an echo of the brain processing sensory information, and thus re-flects connectivity and changes taking place in the developing brain. The first major component of the ERPs is negative peak that appears 100 ms (N100, called simply N1) after the stimulus in adult brains, followed by a positive peak after about 200 ms (P2). Typical N100-P200 complex develops only after 10 years of age. Initially in the first four years the first component is positive (P1 in Fig. here). ERPs result from complex brain processes and are difficult to interpret, but nevertheless are frequently used to analyze various experiments related to perception and language. More sophisticated techniques are based on high-density EEG, with 128 or more electrodes, and algorithms that use signals measured by electrodes on the scalp to reconstruct deeper sources of brain activity and create functional connectome showing the information flow pathways (Hassan et al. 2015). Many other techniques help to understand how brain networks enable complex cognition and intelligence (Barbey, 2018).

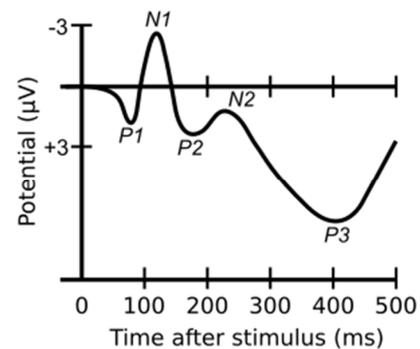

Reading and development of language-related abilities requires precise recognition of phonemes, recognition of graphical symbols (graphemes, logograms), linking this information to form chains of synchronized local brain regions, activating premotor cortex to produce appropriate sounds. Speech input synchronizes activity in the temporal and frontal cortex (premotor, Broca area) forming unique patterns of activation for ordered strings of phonemes, leading to the active subnetworks (resonant states) representing word forms (Grossberg, 2003). Analysis of the N200 feature of event-related potentials shows that phonological processing precedes activations that spread all over the brain by about 90 ms (Pulvermueller, 2003). These extended activations allow for associations and interpretation of the word in a given context, binding phonological representations of symbols with related perceptions and actions, grounding the meaning of the word in a perception/action network. Symbols in the brain can therefore be seen as labels for prototype activations including how they sound like (auditory cortex), how they are pronounced (premotor cortex), what visual, tactile, olfactory, gustatory, emotional and motor associations they have. Such encoding provides easy access to associations, phonological as well as semantic similarities between concepts.

Differences between cognitive abilities may be inferred from event-related auditory potentials immediately after birth. Tests done with reaction to the groups of syllables: bi, di, gi; bae, dae, gae; bu, du, gu; ba, ga, da, performed in the first days of life show that the structure of ERPs (differences in reactions to similar sounds) allows for prediction of emergence of reading disorders 8 years later (Molfese 2008; Molfese et al. 2002). The speed of brain reaction of infants that will have problems in later life is about 200 ms slower. The differences in word recognition of 8-years old reaches 500 ms, and the recognition is not stable, differences between similar sounds are not discriminated with sufficient precision, leading to problems in reading and text comprehension. In longitudinal study Fagan et al. (2009) investigating visual recognition memory (using tests of selective attention to novelty) in babies 6-12 month old found strong correlation with IQ tests when these babies became young adults. These and many other results are in agreement with the idea that prenatal and infant periods are critical in development of brains that facilitate intelligence. Research performed on rats showed that environmental enrichment in early life stages leads to many positive effects, including thicker cerebral cortices, 25% increase of synaptic density and 12-14% increase of glia cells. This increase seems to stay for longer times even when rats are moved to impoverished environment (Simpson & Kelly, 2011). We have not yet learned how to stimulate babies in an optimal way to boost their development at a very early stage, but this is exactly what we would like to do in future.

Segmentation of experience seems to be operating at all levels (Zacks et al. 2010). Practical conclusion for teaching is that all material should be given in appropriate chunks, trying to create as many associations with what is already known as possible (this is called "deep encoding" in psychology). Meanings are stored as activations of associative subnetworks. Different patterns of their activation may influence other areas of the brain that can recognize the type of pattern and process information encoded by such pattern further, associating it with other patterns. For example, hearing a word activates string of phonemes increasing the activity (priming) of all candidate words and non-word combinations (computational models of such phenomena in phonetics are described in Grossberg, 2000; Grossberg, 2003). Polysemic words probably have a single phonological representation that differs only by their semantic extensions. Context priming selects extended subnetwork corresponding to a unique word meaning, while competition and inhibition in the winner-takes-all processes leaves only the most active candidate networks. The precise meaning of a concept is always modified by the context, so explanation of the meaning in a thesaurus can only be approximate. Overlapping patterns of brain activations for subnetworks coding word representations facilitate strong transition probabilities between concepts, activating semantic and phonological associations that easily "come to mind".

Another aspect of brain function important for intelligence and creativity is synchronization and speed. Myelination of long axons (white matter) helps groups of neurons to synchronize quickly and efficiently process information (Singer 1999). The speed of thinking and creation of spontaneous thoughts or images depends on synchronization of distant brain areas. Creativity is one of the most mysterious aspects of the human mind (Duch, 2007, 2007a). Creative brains have trained neural network, encoding dense potentially accessible states and enabling rich association between them. Spontaneous processes arising from blind variations of local excitations of groups of neurons activate various subnetworks forming patterns that depend on the context used for priming. Selective filtering of interesting patterns is based on associations and emotions. This is in agreement with *Blind Variation Selective Retention* (BVSR) psychological model of Campbell, reviewed by Simonton (2010) and presented in computational framework by Duch and Pilichowski (2007).

Learning requires physical changes in the brain, and this is possible due to the neuroplasticity, ability of the brain to change itself (Doidge, 2007). Various processes at the molecular level contribute to changes of synaptic connectivity and activation thresholds of neurons. Pharmacological interventions may influence some of these processes and help in case of severe memory problems. However, good memory is based on strong synchronization of groups of neurons, unique and stable patterns of neural activity. This means strong binding of thoughts, while creativity according to BVSR theory is based on exploration of many associations that is hampered by strong synchronization. Learning and creativity requires intermittent windows of increase neuroplasticity – bursts of excitement, emotions that help to discharge neurotransmitters and neuromodulators – followed by calmer period to stabilize new brain patterns that will be accessible in the future. This is why learning computer games is so quick and why infants learn so quickly. Similar effects may be achieved by direct activation of the brain using transcranial magnetic stimulation (TMS), or direct current stimulation (DCS). There are indications that such stimulation may accelerate the speed of learning and facilitate creativity (Chi & Snyder, 2011).

## EXPERIENTIAL LEARNING STYLES

As with many other processes, like perception of ambiguous figures, memory recall, planning, problem solving, spontaneous creative activity requires 3 steps:
- Preparation of the brain and conscious introduction of the problem.
- Unconscious associations that add new facets providing steps towards solution.
- Conscious recognition of most interesting associations.

Learning requires trained brain network that is primed by the description of the problem. This provides a kind of space for ideas, where new activation patterns and associations are implanted. If the basic concepts have not been learned properly the brain network will not be able to create useful associations. If too many processes in the brain are active priming will not be effective. There is a tradeoff between rigid automatic responses and the ability to make novel associations. Insight processes and involvement of the right hemisphere have been experimentally investigated (Jung-Beeman et al. 2004; Bowden et al. 2005) and interpreted by Duch (2007).

The information that needs to be analyzed must first appear in the working memory. Thus the ability to pay attention, focus on the problem and inhibit irrelevant brain process is important. The problem is easy if relevant features are extracted and associations are quickly formed, as it happens if similar problems have been solved many times. Understanding of basic concepts is equivalent to placing them in the web of associations, using chunks of knowledge that cannot easily be replaced by elaborate reasoning, learning symbol manipulations, that is forming strong associations between different concept representations that automatically and effortlessly lead from one brain state to the other.

Stating the problem by reading, listening or thinking about it puts it into working memory, that is activates (primes) a subset of long-term memory patterns and thanks to sustained attention binds them together. Activation is spread and associated memory elements activated; this may be interpreted as inferences made by specialized processors that can handle bits and pieces of the problem (Baars, 1988). New activation patterns are recognized by the central executive as useful steps towards solution, thus changing the current state of the problem; this cycle is repeated until a solution is found or an impasse is reached. Final solution is a series of associations that lead from the initial brain state – problem statement – to the final state, representing problem solution.

This view leads to unified approach that should be the basis of learning: prepare the brain, introduce the problem, wait for the solution. This process may be done effortlessly but requires ability to focus and to prepare the brain for learning. Preparation of the brain may be done using mental relaxation response techniques (Benson, 2001), or physical exercises that require focusing on bodily sensations. The use of neurofeedback techniques has also been quite successful (Gruzelier, 2009). However, memorizing basic facts needs to be done first to create necessary basis for the space of concepts that is necessary for solving problems in a given domain. The *Core Knowledge Foundation* tries to teach such basic concepts that should be useful to everyone, starting from preschools, with programs in mathematics, number sense, and orientation in time and space.

In many cases the brain may take intuitive decisions evaluating complex similarity patterns – activation patterns of cortical networks in posterior sensory and associative cortex will automatically be perceived by the working memory executive frontal lobe areas as similar, because information carried over such long distances in the brain is not so precise. The number of logical rules required to justify some decisions based on intuition may be impractically large. Explanation of intuition is thus rather simple (Duch 2005, 2007).

The individualized learning styles have been discussed in learning sciences since 1970. Several theories have been proposed. Perhaps most popular is based on three learning modalities (Barbe & Swassing, 1979): visual, auditory and kinesthetic (hence the acronym VAK). However, this idea has been severely criticized by many experts and there is no indication that it benefits education; it has been even called a "persistent myth" (Kirschner, 2017). The existence of learning styles (or learning preferences) is hotly debated, and a large number of papers on this subject are published each year.

Among all learning styles theories developed so far the experiential learning approach presented in David Kolb's book (2015, 1st ed. 1984) has perhaps the best empirical support. Kolb views the learning processes along two dimensions: preferred mode of perception – from concrete to abstract – and preferred mode of action – from individual experimentation to reflective observation (Pashler et al. 2008). This division leads to 4 extreme types of learners: divergers (concrete, reflective), assimilators (abstract, reflective), convergers (abstract, active), and accommodators (concrete, active). The *Learning Styles Inventory* is a tool used to determine individual type, and is distributed commercially. Reality is not black and white, therefore these dimensions should be viewed as a continuum.

Some neuropsychologists could not see relations of these ideas to the neuroscience and therefore have criticized learning styles as baseless. In the second edition of his book Kolb added chapters relating his ideas to brain functions, as described by J.E. Zull (2011). However, neither Zull nor Kolb mentions connectomes. Our current understanding of information flow in brains is very relevant to different styles and types of learners.

Let's distinguish 3 types of brain activity:

1) sensory level S, with strong synchronization of groups of neurons that are involved in processing auditory, visual and other sensory information (mostly occipital, superior temporal and somatosensory cortex);
2) central level C, with abstract concepts that have no sensory components and thus engage neurons mostly in association cortices (parietal, temporal and prefrontal lobes), and
3) motor cortex M, that involves activations of motor imagery and physical action (mostly frontal cortex, basal ganglia and cerebellum).

Of course all brain areas are coupled, so division into 3 subsystems provides only the roughest description, but in individual experience there are moments when focus on sensory information, abstract thinking or motor action clearly dominate. Concentration may shield against distraction reducing influence of background processes on global brain state. At first approximation individual differences can be linked to flow of information between 3 subsystems responsible for sensory perception, memory and abstract thinking, and control of motor action. In case of autism local overconnectivity within each of these regions, and weak connectivity between them is well documented. Functional magnetic resonance study of connec-tomes of autistic people showed group differences between brain states that primarily involved sensory and motor networks, and those involving higher-order cognition networks (Mash et al. 2019). Although this topic requires much more research correlation between individual preferences in learning and involvement of large brain networks should be expected.

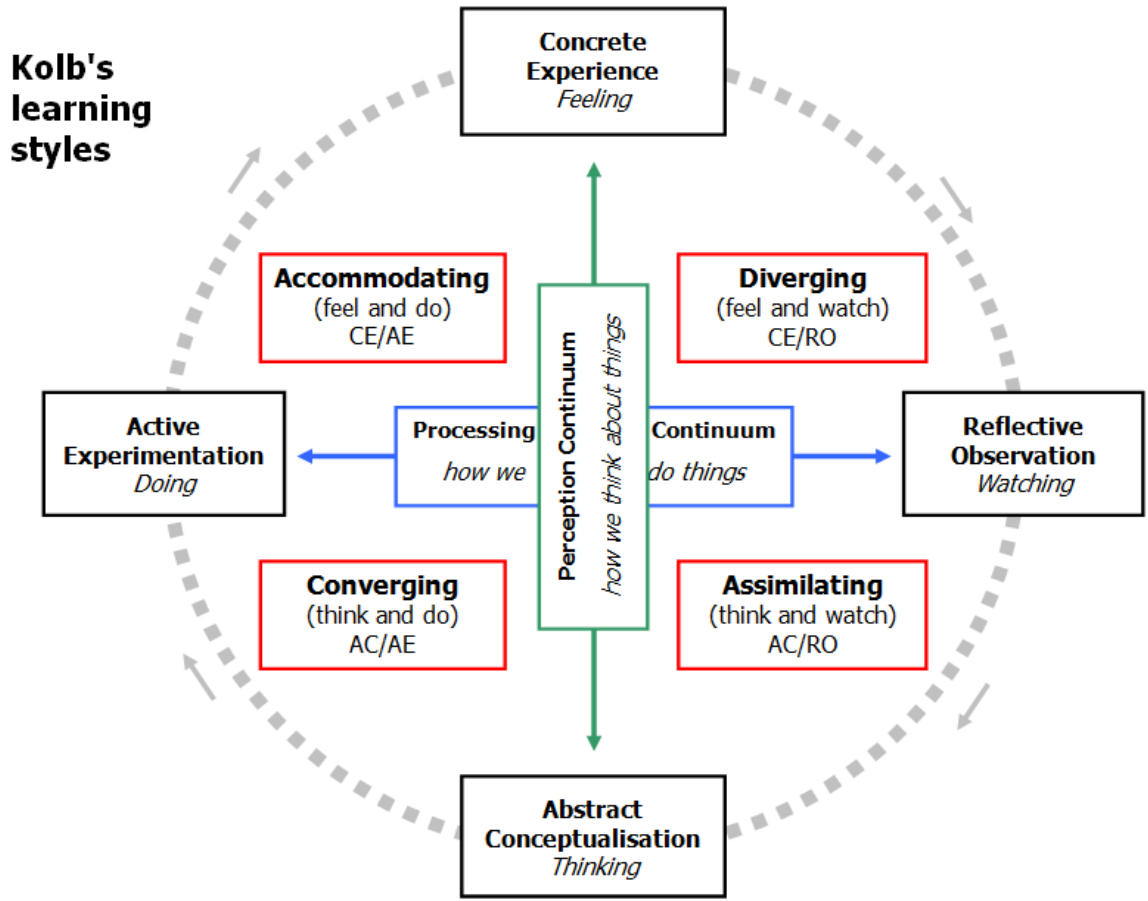

Fig. 1. Graphical summary of two dimensions and 4 extreme characteristics of learning styles according to Kolb (2015).

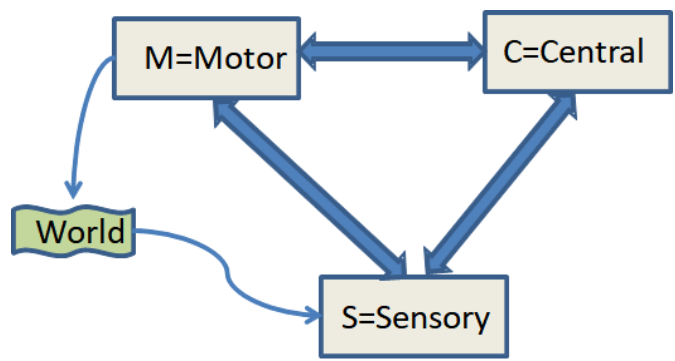

Fig. 2. In optimal case all brain areas are integrated and information flows between sensory, motor and higher-order associative networks.

The first dimension proposed by Kolb involves observation, watching vs. doing. Watching may be linked to the dominance of sensory processes, strong coupling of sensory S⇔S activations, with weaker coupling between abstract central processes C⇔C. This situation is quite common during demanding action games, including ping-pong and point-and-shoot computer games, when brain power is used for sensorimotor action, with no time for reflection. Sensory and motor activity are strongly coupled, learning games leads to changes of both sensory and motor networks, and observing results of motor action on the environment (in Fig. 3 this is external loop) also the coupling between both areas. Brain activity is dominated by synchronization between motor and sensory processes M⇔S and internal S⇔S , M⇔M activity (Fig. 3).

John Dewey has popularized the idea of learning by doing (see Kolb 2015 for history of experiential learning). In complex cases involvement of central processes is also necessary to build a model of the situation, but many crafts can be learned by watching and actively experimenting. Learning, as Kolb writes, is indeed a continuous process grounded in experience. Parietal areas are involved in representation of even simple information, such as speed of movement (Hamano et al, 2020), so some information flows from S⇒C, but broader C subnetworks are not activated. Brain dynamics of some autistic people may be dominated by processes in the motor networks M⇔M, with repetitive movement or echolalia.

Attention requires synchronization of groups of neurons. During sensory perception neurons send about 40 spikes per second (40 Hz), and during anticipatory attention about 20 Hz (Kamiński et al, 2012). Ability to focus sustained attention on sensory stimuli for longer time shows high variance among people. In the case of some autistic people strong synchronization of sensory networks makes the shifts of attention between stimuli quite difficult, as it has been demonstrated in numerous experiments (Bristol et al. 2020). This dimension has thus 3 extremes states, in which sensory, motor, or sensor-motor activity dominates.

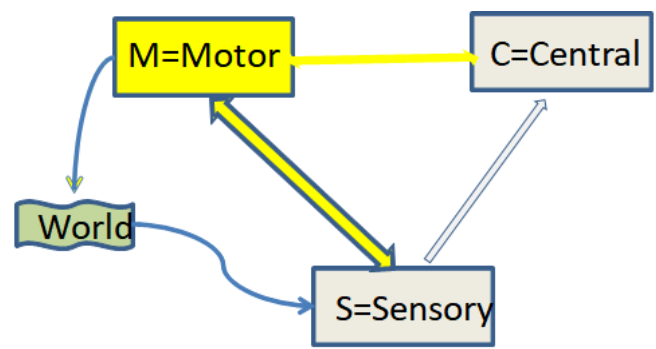

Fig. 3. Strong coupling within sensory networks, and sensory-motor connection, with weak coupling with higher-order associative networks.

The second dimension involves concrete and abstract processes, experiencing and feeling vs. thinking and imaging. It can be understood as gradual domination of central processes, decrease of sensory and motor input in the content of global brain dynamics that is responsible for inner discourse. It may result in abstract reasoning, directed by prefrontal executive networks, or it may be self-directed personal reflections, that involves the default mode network areas in parietal and prefrontal lobes (medial prefrontal cortex, posterior cingulate cortex and angular gyrus in parietal lobe). Anterior cingulate cortex may also be involved in meta-cognitive processes giving us the sense of self.

If the top-down links are weak the central C⇔C processes will dominate, no vivid sensory imagery should follow thoughts, but abstract thinking may be more efficient. This type of thinking may characterize mathematicians, logicians, theoretical physicist, theologians and philosophers who prefer to think about abstract ideas. Stronger bottom-up links S⇒C will activate association cortex without strong activity at the S⇔S level, leading to abstract thoughts. In case of people with Asperger syndrome, a milder version of autism, the lack of understanding of metaphoric language indicates that reference to physical objects also activates sensory cortices strongly, although perception does not constrain brain activity so strongly as in the cases of autism.

Strong top-down links between association cortex and visual and auditory cortices will lead to vivid imagery dominated by concrete sensory experience. In extreme cases this is present in autistic people, who have very vivid and detailed imagery, but little abilities to generalize experience or relate it on a more abstract level to general knowledge. Activation of limbic structures (strongly connected with medial prefrontal cortex that regulates emotional responses) may add emotional content to sensory experience.

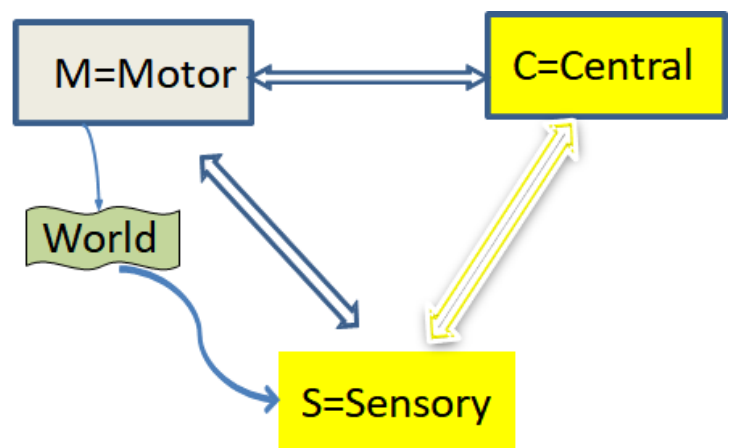

Fig. 4. Strong coupling within sensory networks, and sensory-motor connection, with weak coupling with higher-order associative networks.

This general scheme helps to interpret 4 types of learning styles proposed by Kolb in terms of simplified connectome between 3 major brain regions.

**Assimilators** think and watch: they are prone to abstract thinking, reflective observation, inductive reasoning. This may be due to the strong connections S⇒C and activity within C⇔C, with weaker connections from S⇒M and C⇒M. Brain processes that control behavior are in this case dominated by reflective observation, inductive reasoning and abstract conceptualization. This is typical of people who spend their time thinking, philosophers, mathematicians, lawyers, computer designers, linguists, or theoretical scientists. If the top-down connections C⇒S are rather weak imagery will tend to be abstract, without significant sensory component.

**Convergers** combine abstract conceptualization with active experimentation, using deductive reasoning in problem solving. Strong C⇒M and C⇔C flow of activity leads to converger style. Experimental scientists, engineers, economists, corporate lawyers, medical professions are in this group. Thinking and building models of situation is combined with action. Imagery is as important as are the results of physical actions.

**Divergers** focus on concrete experience S⇔S, therefore strong C⇔S influence is expected, with C⇔C activity facilitating reflective observation, strong sensory imagery, but weak influence on motor activity that is mainly to increase sensory experience. In this category we may find administrators, coaches, social advisors, influencers, writers, journalists, political and culture experts, civil lawyers, computer programmers working on relatively simple assignments, who frequently use trial and error strategies.

**Accommodators** have balanced communication between sensory, motor and central areas and thus combine concrete experience with active experimentation supported by central processes S⇔C⇔M. However, they are not trapped by abstract thoughts, so higher-order networks C⇔C are not too strong, and they are also not trapped by sensory experience S⇔S, acting and observing. We can find here managers, experts who do experimental work, mechanics and mechanical engineers, electronics repair, plumbers, teachers.

It may be hard to link vocations with learning styles, as in many professions there are different subtypes of experts. For example, in physics or chemistry we have a whole spectrum of demands, from engineering who construct equipment, experimental chemists and physicists performing tedious measurements, computational physicist who construct models and spent most time on programming, to applied mathematicians that work on highly abstract and speculative theories.

Learning styles can thus be linked to communication between major areas in the brain and to the dynamics of brain processes in major subnetworks. Information flow is not symmetric, so even this simple model of connectome based on 3 areas leads to many possibilities. For example, sensory imagery is very useful to artists and musicians, and it requires strong top-down connections, i.e. C⇒S. On the other hand abstract thinking may be easier if such connections are weak, and more energy is spent on the internal C⇔C activity. How many situations can be distinguished? Suppose that we have a threshold for functional connectivity measure, with higher values signifying a strong functional connections (flow of information) and lower weak connections. Two areas are always at least weakly mutually connected; there may be a strong connection in one of the directions, or in both directions. That makes 4 cases. We have 3 pairs, (S,C), (C,M) and (S,M), each may be in 4 states, so there is 12 cases. In each of these cases we may have strong local activity in one of the 3 regions, increasing the number of different states to 3*12, or in two regions (another 3*12) or in all 3 regions (another 12 states). The total number of combinations will be 7*12=84. In addition brain state is dynamic, depending on the physiological state of the organism, emotional arousal, so there will be transitions between these states. Strong local activity at S may decrease or increase information flow to C and M, depending on the kind of sensory stimuli.

Suppose that we could collect a lot of information about functional connections between brain regions in different situations. Functional MRI provides connectome data only in simple experiments (due to physical constraints in the scanner) and does not tell us much about direction of the information flow. Wireless EEG techniques may provide more information in complex situations, at least about processes in neocortex. Will such data show stable clusters of activity that could be used to characterize cognition, based on objective measurements, in contrast to Myers-Briggs or other personality inventories? Psychology strives to describe reality in simple terms, but what if psychological constructs used to create models are not able to describe real brain dynamic processes behind cognitive behavior (Duch, 2018).

## SUMMARY AND CONCLUSIONS

An overview of various factors that contribute to learning has been presented, with a new analysis of learning styles inventory based on simplified connectome, linking the peripheral (sensory), central and motor brain areas. Most research programs related to phenomics at different levels: genetics, epigenetics, signal pathways, neurons and their networks, are at pre-sent aimed at serious mental disease rather than understanding or improving learning. It is clear that we are at the beginning of a long way towards neurocognitive phenomics that will take a long time and will require a lot of effort to develop. Analysis presented here shows that even simple models of information flow in the brain have large number of variants and may not be stable, influence by external environment and internal physiological changes. Such approach is in danger of being too flexible, able to explain everything at least qualitatively. All kinds of learning preferences analyzed psychologist and experts in learning sciences may be linked to one of brain states described in the previous section. On the other hand brains are the most complex systems in the known universe, so it is not clear that description of behavior may be simplified in a useful way using models based on psychological constructs.

Many important ideas related to education, such as the role of emotions, training of motivation, exploration, strong will, depletion of willpower, goals setting, have not even been mentioned, although they are very important. Many ideas that education has been experimenting with are accommodated in a natural way in neurocognitive phenomics. Understanding brain processes explains some aspects of learning, creativity, automatization of skills, insight and many other processes that were quite mysterious not so long time ago. This view opens many possibilities for technological support of development of full human potential, starting from the prenatal period and the infancy.

Until the beginning of this century the field of artificial intelligence was mainly concerned with verbalized, symbolic description of knowledge, ignoring development in machine learning, neural networks and neuroscience. Successes in deep neural networks and other deep learning algorithms have completely changed the main direction of AI research, which tries now to incorporate all related fields. Currently educational neuroscience develops in parallel to the learning sciences (Azilawati, Henik, and Hale, 2019) and there seems to be a little overlap between these two fields. However, explanations at psychological, neural and other phenomic levels complement each other (Howard-Jones et al. 2016), and at least rough understanding of neuroplasticity and neural information processing in the brain should be included in the foundations of learning sciences.

**Acknowledgements**. This research was supported by the Polish National Science Center, grants NCN, UMO-2013/08/W/HS6/00333, and UMO-2016/20/W/NZ4/00354.

<div align="center">

BIOGRAPHY

</div>

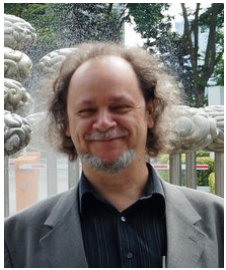 **Włodzisław (Wlodek) Duch** started his scientific career from theoretical physics, then moved computational intelligence (CI) methods developing meta-learning schemes that automatically discover the best model for a given data and is working now on development of neurocognitive informatics, algorithms inspired by cognitive functions, information flow in the brain, learning and neuroplasticity, understanding attention deficit disorders in autism and other diseases, infant learning, toys that facilitate mental development, creativity, intuition, insight and mental imagery. He worked as a Visiting Professor in Germany, Japan, Singapore and USA, and currently serves as the Vice-Rector for Research and ICT Infrastructure at Nicolaus Copernicus University in Toruń. Search for "W. Duch" to see his full CV.

These results indicate that the originality of CMDT is associated with (a) greater activation of the ventral attention system, which is involved in reorienting attention and (b) reduced task-induced deactivation of the default mode network, which is indicative of alterations in attentional reallocation, and (c) cognitive correlates of originality of CMDT and revealed sex differences in these associations.